# How to Achieve Justice through Big Data: A Study Based on Credit Card Cases in Beijing

Authors: ChengTo Lin (1), Chung Han Tsai(1), Baowen Zhang (1), Qingyue Deng(2), Yunhui Zhao(2), Zhijia Song(2)
((1) Renmin University of China, (2) Shandong University)

# Abstract

With the development of intelligence, the combination of big data and judicial practice has become a hot research topic. There are fewer studies on credit card contract disputes related to big data, which makes it difficult to respond to the trend of Big data era. This paper uses the data source of credit card disputes related to the protection of personal information in Beijing from 2022 to 2024， introduces a social network analysis methodology to analyse validated judgements and creates a database. Through the analysis of the relationship network indicators of applicable laws and regulations, it can clearly show the laws and regulations applicable to this type of dispute in China in terms of substantive and procedural matters. This paper provides a methodology for combining big data with judicial trials and assisting judges in trials. The effective use of social networks will help fill the gap in the application of big data in judicial trials. The result and discussion of the paper is also of practical significance for improving the efficiency of judicial adjudication, constructing a database for searching classified cases, constructing a holistic system for digital courts, and finally promoting justice.

# Introduction

## Research background

The escalating strain on China's judicial resources, with a caseload growth rate outpacing the expansion of judicial personnel, has resulted in a critical issue of 'Too few judges handling too many cases' (Su L.,2010). How to improve the efficiency and quality of the hearing of cases, allowing each case to be fully and fairly adjudicated has become a judicial challenge that China is now facing.

In 2022, the average number of cases per capita of judges in the primary people's court of China was 274, and in nine of these provinces, the average number of cases per capita of judges was more than 300, with the highest number being more than 400 (Cheng JH.,2022). The double challenge of enormous case-handling pressure and increasingly tight judicial resources must be addressed through the digital transformation of the courts, which can optimise the efficiency of litigation, improve the quality of trials, ease the pressure on judges as well as provide more intellectual aid and support to them (Jia Y.,2024).

Digital courts in China are enhancing judicial justice and efficiency through big data analytics. At the 2024 National People's Congress (NPC) and the Chinese People's Political Consultative Conference (CPPCC), Zhang Jun, President of the Supreme People's Court, pointed out that we should strive to realise that more than 3,500 courts across our country will be able to handle cases on one network (Zhang J.,2024). By using judicial big data, the court would complete the transformation from 'traditional manual trial' to 'data scientific analysis', thus solving the typical problems faced by traditional court trials, such as inconsistent application of laws, delayed and slow procedures, and judicial corruption, so that truly achieve judicial justice.

A variety of local courts are also actively exploring digital courts. In Beijing, for example, in the first half of 2022, online handling of cases of the city's courts accounted for 82.9% of the number of first-instance civil and commercial cases received during the same period (Zhao Y., 2024). They explored the application of digital and big data to the trial process, becoming one of the earlier courts in China to explore the digital court. In 2023, the High People's Court of Shanghai Municipality took the lead in proposing and comprehensively implementing the construction of digital courts. At present, the construction system of the digital court, the operation mode, and the promotion path have been shaped and established. The efficacy of the application scenarios has gradually appeared. (Putuo District of Shanghai Municipality, 2024)

The meaning of this research lies in its innovative textual measurement of judgement content, and then the use of social analysis networks (SNA) to form a judicial information network, extracting paradigms from its results and applying them to the basic trial database of the digital court. In theory, it analyses the advantages and shortcomings of this model in solving the current problems of digital courts and points out the direction of future optimization.

In practice, it provides possible solutions to the challenges currently faced by digital courts, such as the imperfection of the trial quality inspection system and the insufficient depth of judicial data mining. By linking judgements of similar cases, firstly, the model can help build a more accurate case base of dispute typology and judicial data network, helping judges to quickly find the relevant laws and to summarise the laws and experiences from the comparison of cases. Secondly, it provides an effective tool for monitoring trials and improving the quality of case hearings, which contributes to the realization of justice in terms of improving judicial efficiency and strengthening judicial oversight.

## The characteristics of credit card disputes and their relevance to the digital court application

First, the number of such disputes is high, and there is an urgent need for digital technology to alleviate the dilemma of “Too few judges handling too many cases.” in China. In the judicial practice of credit card disputes, there are unavoidable overlapping or conflicting parts of the user’s personal information and credit card collection data, which may easily give rise to certain legal issues of personal information infringement. From 2018 to 2021, there were over 1,000 cases of personal information protection disputes in credit card cases nationwide each year, and the trend was growing year by year. This means that the application of digital technology to conduct intelligent, batch trials can reduce the work pressure of judges' manual trials and alleviate the current situation of "Too few judges handling too many cases" adjudication in China.

Second, the application of standard terms and summary procedures to credit card disputes contributes to digital batch processing and helps trials proceed efficiently.

At the same time, the research found that, over time, credit card contract disputes in the judicial process more and more application of summary procedures, January to August in 2022, Xicheng District of Beijing Municipality concluded 51,013 civil and commercial cases of trial of the first instance. Among them, 27,252 cases were closed by applying small claims and summary procedures (accounting for 53.42%) reflecting the judicial hotspots in recent years (Xicheng District of Beijing Municipality, 2024).

This paper uses PKU Law Database as a source of judgement information to search for two kinds of individual credit card dispute cases (2020 and beyond, from Beijing and all around the country)[1] for preliminary screening query, the total number of cases applying summary proceedings line graph as follows figure 1, 2.

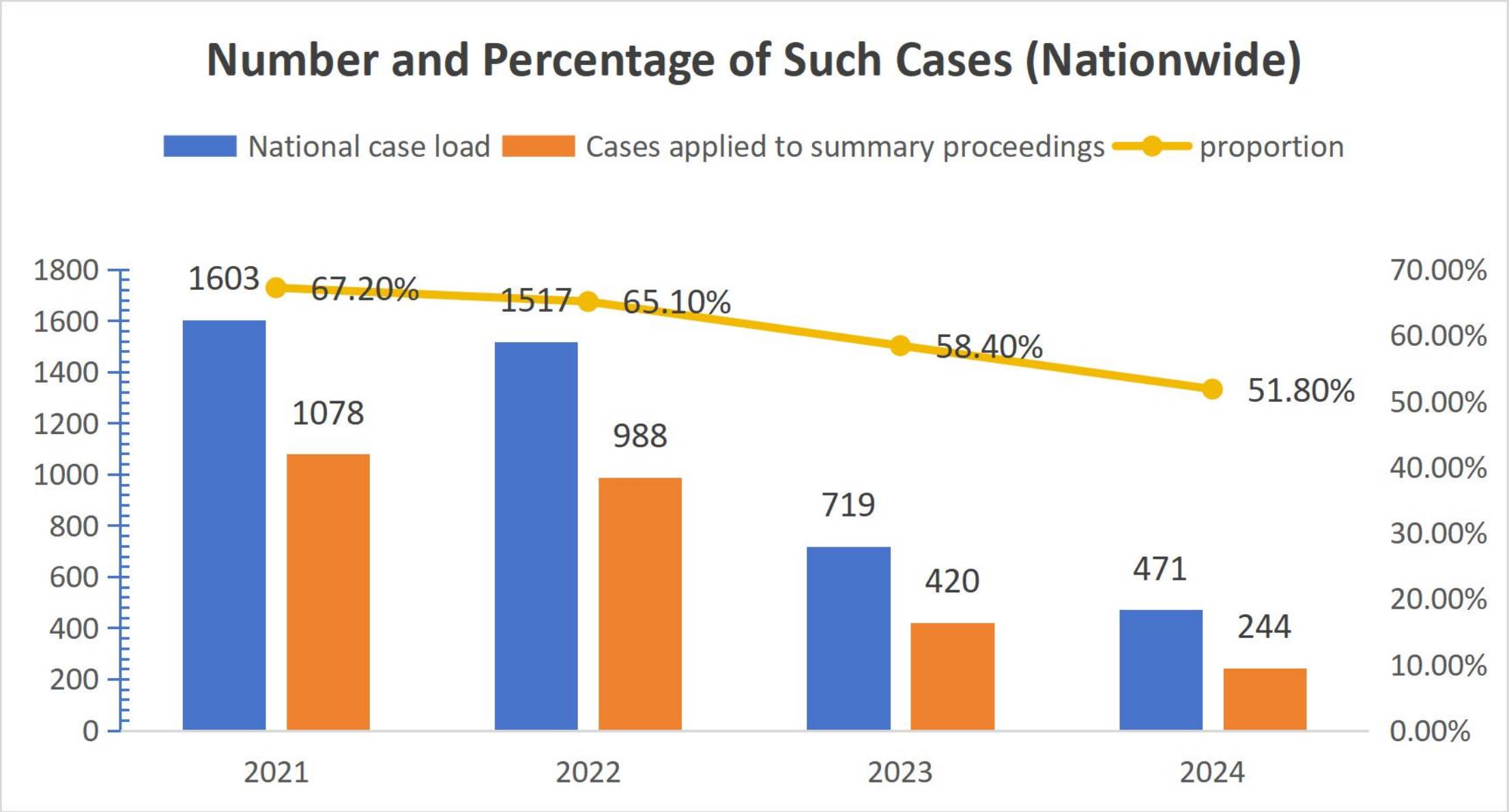

Figure 1

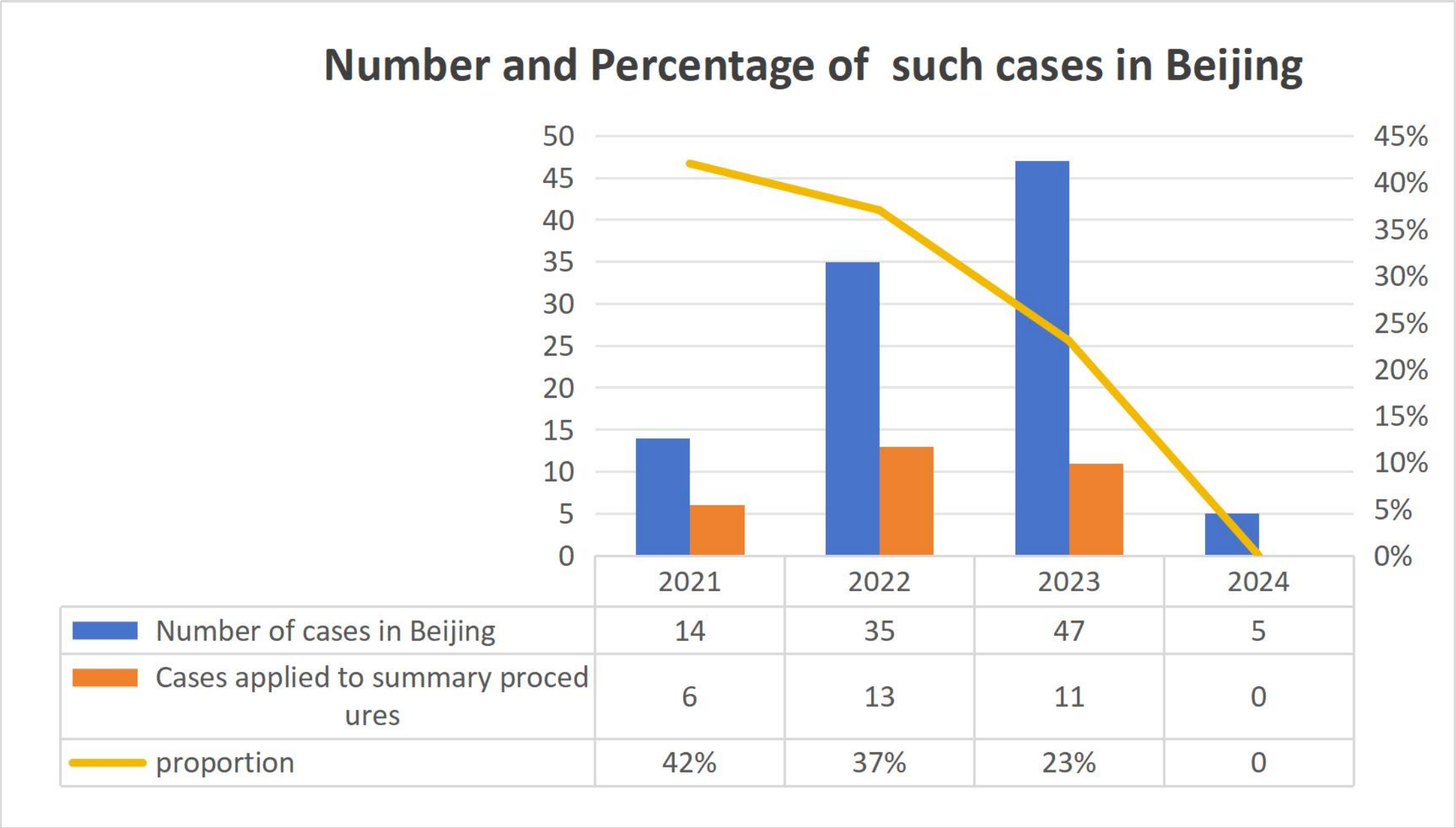

| | 2021 | 2022 | 2023 | 2024 |
|---|---|---|---|---|
| Number of cases in Beijing | 14 | 35 | 47 | 5 |
| Cases applied to summary procedures | 6 | 13 | 11 | 0 |
| proportion | 42% | 37% | 23% | 0 |

Figure 2

From the chart above, the research found that the number of personal information protection disputes in credit card cases in which summary proceedings were applied remained at around 50-60% of the total number nationwide.

According to Civil Procedure Law of the People's Republic of China: Where a primary people's court and its detached tribunals try simple civil cases with clear facts, unambiguous rights and obligations and minor disputes, the provisions of summary procedure shall apply[2].

Therefore, personal information protection disputes in credit card cases are conducive to the application of big data for batch trials in digital courts because of the following three main features: Firstly, regarding contractual documents, they apply standard terms, with clear rights and obligations relationships and obvious homogenisation features. Secondly, in terms of the content of their disputes, the facts are clear and the difficulty of resolving them is low, which could apply to summary procedures. Therefore, this paper takes personal information protection disputes in credit card cases as an example to research the application of social network analysis (SNA) in digital court intelligent trial system.

## The current state of research on the Smart Trial System of

## Digital Courts

It has been found that, in judicial practice, the Digital Court Intelligence Trial System has been used to a certain extent in preliminary applications, which are mainly divided into the following three categories: the first category is auxiliary adjudication tools for case topics. It can screen the elements of a specific theme case, help the judge quickly catch the focus of the case and evidence, improve trial efficiency. For example, courts in Shanghai have developed and applied the ‘Shareholder Right to Know Dispute Elemental Trial Aid Model’ and the ‘Property Search and Control Alerts for Cases of Enforcement Termination (Accumulation Fund)’ data model, and so on (Chen X., Chen L., Wu Z., 2024). The second type is a comprehensive tool for the entire trial process, assisting manual work in different parts of the trial procedure. For example, the case evaluation system developed and launched by the Xicheng District of Beijing Municipality: ‘The system has the functions of randomly dividing cases, automatically associating electronic file archives and labelling case types.’ (Fan C, Hu YY, Qi PY, 2024). Another example is the Intelligent Procedure Control Robot it developed and applied, ‘which can automatically operate through big data analysis to collect information on parties, initiate the execution of property search and control procedures, and help execute the results of the adjudication.’ (Ibid). The third category is supervision tools for the trial process, which regulate litigation procedures and correct procedural irregularities. The third category is supervision tools for the trial process, which regulate litigation procedures and correct procedural irregularities. For example, the Shanghai Municipal Court has established ‘prompting for not lifting restrictive measures after execution’, ‘prompting for defective service in announced cases’, and ‘prompting for refunding litigation fees to successful parties’ (Jia Y., 2024). The research study concluded that digital court trial practice now has about three types of application paths (as shown above).

In academic research, most of the practical research on the legal application analysis and decision-making module in the digital court intelligent trial system focus on theories and lack the support of corresponding visualisation models: e.g., Wang in 2024 proposed to abstract the relativity between the constituent elements of the case facts and the legal facts in the digitally-assisted case management to achieve the dismantling and digitisation of the legal norms (Wang SX., 2024); Chen Y. and Sun Y. in 2023 proposed to extract legal text information in judicial cases by introducing generative artificial intelligence technology based on pre-trained language models (GPT) (Chen Y., Sun Y., 2023), etc.

From the application mode of adjudication cases, in Germany and other civil law countries, the application of jurisprudence mode is 'analogy - induction – deduction' (Ibid). China belongs to statute law countries, similar to the analogy process of that to build the legal thinking model of case search: Firstly, searching for precedents that are similar to pending cases. Secondly, induct general adjudication features from precedents that transcend case-by-case judgements to explore the judicial logic present therein and construct a model of adjudication rules for similar cases. Thirdly, use deductive reasoning to connect with the facts of pending cases to assist the judicial trial process.

In summary, in terms of application, the intelligent trial systems of digital courts are mainly divided into the above three major categories, and there are fewer auxiliary trial models for exploring the application of law in adjudication. In terms of academics, most of the research also focuses only on developing directional theoretical research.

This paper argues that the judicial information database established by the social network analysis method based on text metrics can aid digital courts in many aspects of adjudication, such as case typing, intelligent application of law, and construction of informatised databases, and so on. Therefore, in order to explore the application possibilities of the database built based on this method in the intelligent trial system of the digital court as well as to help the court promote the development of auxiliary adjudication models in the process of constructing the digital court, this paper carries out the trial construction and analysis of the model.

## Research purpose

This paper would mainly explore three major issues in the application of digital courts: First, analysing credit card data disputes to streamline digital court case classification. Second, analysing legal applications in personal information disputes of credit card cases aims to establish an adjudication framework for consistent judgements. Third, addressing data fragmentation in judicial databases involves creating a clear, searchable information database, fostering a comprehensive system, and applying social network analysis to solve social issues and enhance judicial justice.

The overall structure of this paper is shown in Figure 3:

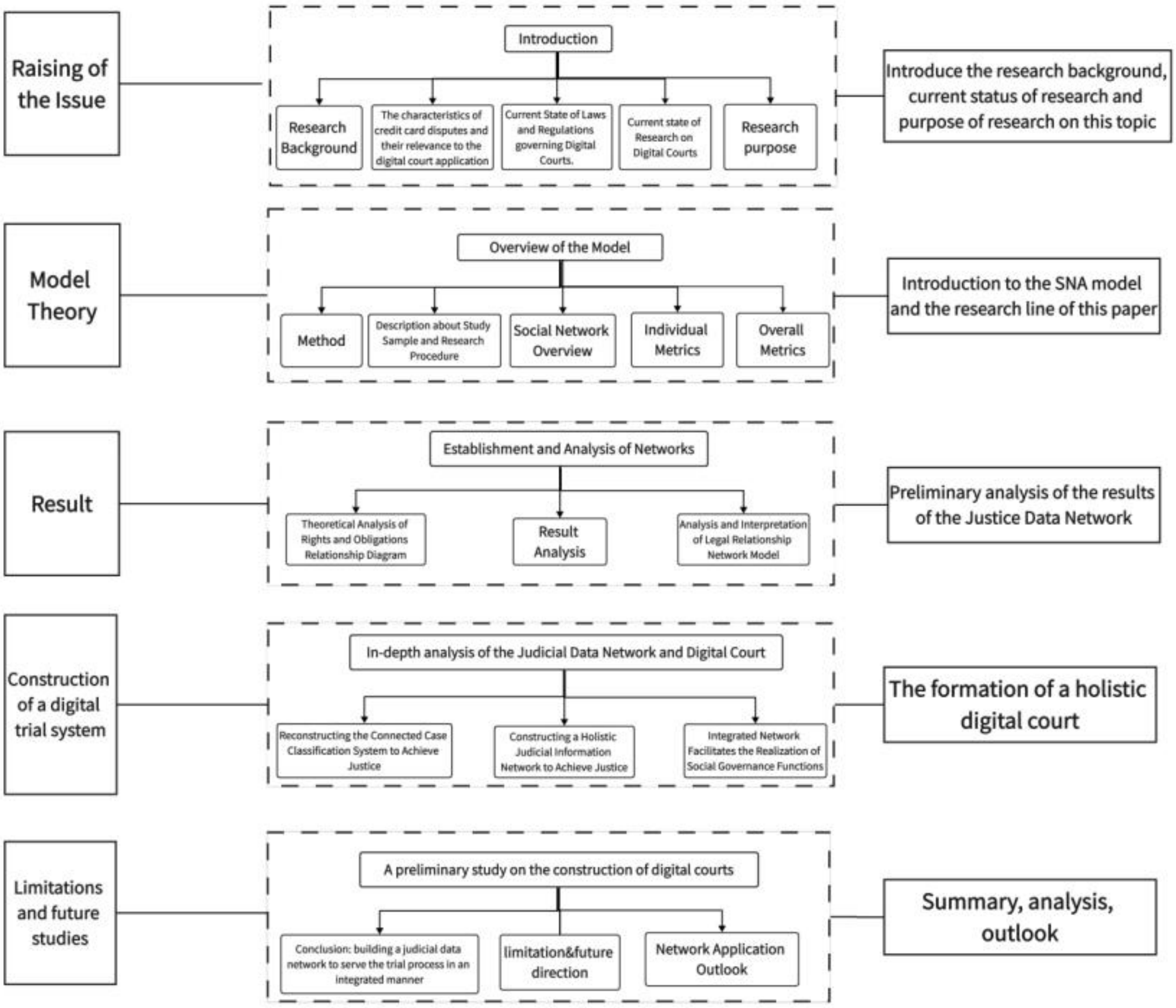

Figure 3

# Method

## Research Method Introduction

The paper aims to construct a citation network between different laws and regulations, using quantitative analysis to identify the most frequently cited combinations of laws and regulations. By doing so, it seeks to summarize the judicial logic involving personal information protection in credit card disputes, thereby facilitating further discussions. Traditional tools such as literature networks and word clouds are not well-suited to displaying the co-citation of different laws, making it difficult to construct a network of legal and regulatory citations. Therefore, this paper employs the method of Social Network Analysis (SNA) to better achieve its research objectives.

Social network analysis (SNA) traces its roots back to the 1930s and 1940s. The concept of a social network refers to a relatively stable system formed by social relationships among certain individuals (Wellman B, 1979). SNA encompasses a set of norms and methods for examining the structure and attributes of social relationship systems. Its primary focus is on analyzing the structure and attributes of relationships among different social units (individuals, groups, or societies). By integrating SNA with empirical analysis, this paper merges overall descriptive indicators with dynamic research, enhancing the persuasiveness of the findings.

In recent years, scholars have increasingly utilized social relationship networks to establish community networks and analyse the influence and correlation characteristics among individuals within these networks. For instance, Isidro Maya Jariego explores the use of stakeholder network analysis to improve the effectiveness of participatory water governance, emphasizing its role in promoting decentralized decision-making and consensual management (Isidro M.J., 2024). Furthermore, Doe J, Smith J, and Brown M use social network analysis to explore collaboration patterns in neuroscience, revealing key players and highlighting the importance of interdisciplinary collaboration for research impact (Doe, J., Smith, J., Brown, M., 2023). Based on the current research, this paper finds that the social network analysis method is widely used in the academic field to study the interaction effects of multiple factors within a certain spatial range, thereby discovering the correlation relationship and studying the structural characteristics of the overall network based on this relationship.

With the advancement of model applications, researchers have gradually applied the social network analysis method to the field of law. The combination of social network analysis and legal research enables the analysis of the pathways through which law and society interact, or to further analyse the implementation of specific laws based on the application of law. The exploration of the research application of social network analysis from a legal perspective is primarily concentrated in the past decade. For example, Masías compared and analysed the effectiveness of using social network analysis to model criminal behavior for legal results, “criminal trial results,” through methods such as logistic regression, Bayesian networks, and random forests, and concluded that social network analysis is the most effective method for analyzing the relationship between criminal law and punishment (Masías V H, Valle M, Morselli C, et al., 2016). Similarly, Coupette C, Beckedorf J, Hartung D, Bommarito M and Katz, D.M present a comprehensive framework for analyzing legal documents as multi-dimensional, dynamic document networks, using network analysis tools to study the evolution and structure of statutes and regulations in the United States and Germany over 20 years (Coupette, C., Beckedorf, J., Hartung, D., Bommarito, M., Katz, D. M., 2021). Legal research with social network analysis can better discover the interconnection between different regulations and policies in the same field. The paper believes that by establishing a database for analyzing the citation of relevant laws and regulations through social relationship networks, it can summarize the implementation of the law, such as solving the problem of lack of empirical effectiveness in current legal research based on big data of legal practice judgements.

The paper believes that social network analysis can study the implementation of law in specific litigation fields. Based on this, it can assist in judging judicial trial logic and play a role in the construction of digital court case retrieval databases, the formation and improvement of specialized dispute classification databases. For credit card contract disputes, which present typological features in the case, constructing a social network can clearly understand the overall judgement logic, reduce the time and manpower costs in the adjudication of similar cases, and provide a feasible path for issuing judicial opinions.

## Description about Study Sample and Research Procedure

This study employs Peking University's Legal Expert database as the primary source of legal documents, focusing on credit card contract disputes. The search parameters were set to include cases adjudicated in Beijing, with judgement dates spanning from January 1, 2022, to September 30, 2024. This yielded a dataset of 49 judicial decisions, from which one duplicate was excluded, leaving 48 unique judgements for social network analysis.

Utilizing the insights gleaned from the social network model, the article constructed a citation network of legal provisions based on court trial practices. This network was designed to investigate the application and significance of legal provisions within the context of specific judicial decisions. The analysis aimed to discern the applicability and importance of pertinent legal articles in judicial adjudication, also to characterize the overall features of the citation network. Such an approach offers a comprehensive understanding of the inherent logic underpinning credit card contract dispute resolutions over the past three years, facilitating a nuanced analysis of the broader trends.

The methodology is delineated as follows: initially, a database was compiled from the 48 selected legal documents post-initial screening; subsequently, UCINET software was employed to generate a matrix, thereby creating a network diagram of legal provision citations, with various metrics calculated; finally, the analyses and conclusions were drawn based on these metrics, adhering to the rigorous standards of academic publication.

## Social Network Overview

In the realm of social network analysis, networks are fundamentally structured by the interconnections among various nodes. Methodological approaches to measuring these networks encompass the evaluation of ties, the analysis of individual nodes, and the assessment of network-wide properties. For individual nodes, key metrics include degree and betweenness centrality, while network-level properties are characterized by dimensions such as size and density. This study concentrates on these indicators to dissect the constructed network, where nodes represent the legal provisions referenced in judicial rulings, and ties are defined by the concurrent appearance of different legal provisions within a single judgement. The analytical framework will be applied to a database compiled from a meticulously selected and refined set of 48 judicial documents.

## Individual Metrics

1. The degree of a node refers to the number of nodes directly connected to a particular node. The higher the degree of a node, the more central its position in the entire network, thereby possessing a higher level of dominance. The degree is an important structural characteristic of individuals within the network. In the legal provision citation network constructed in this study, if a legal provision possesses the highest degree, it is considered to be in a central position, further indicating that the provision is cited more frequently in specific judicial practices, thus occupies a core role in the handling of credit card contract dispute cases.

2. Betweenness centrality measures the extent to which a node in a social relational network is located "between" other points in the graph. The higher the betweenness centrality of a legal provision, the stronger its control over other legal provisions. Concurrently, this implies that the contractual content risks associated with the legal provision constitute the “critical intermediary risk” of the dispute network. By controlling this risk, one can manage the occurrence of the entire risk network. The calculation formula for betweenness centrality is as follows:

$$C_B(n_i) = \frac{\sum_{j<k} g_{jk}(n_i)}{g_{jk}}$$

In the above formula, CB (ni) represents the betweenness centrality, gjk is the number of shortest paths between nodes j and k, gjk (ni) is the number of shortest paths from j to k that pass through node i, and g is the total number of nodes in the network.

## Overall Metrics

（Ⅰ）Size refers to the number of nodes contained within a social network, and the size of the network influences the relationships between individual nodes. In this study, size corresponds to the total number of legal provisions that appear in the sample. A larger network size indicates a greater number of cited legal provisions, suggesting a more diverse range of risk types.

（Ⅱ）Density refers to the degree of connection between nodes within a social network. A higher density indicates a closer relationship between legal provisions, meaning that two or more legal provisions are more frequently cited together in the same judgement. The calculation formula for density is as follows:

$$\mathrm{D} = \frac{2L}{g(g-1)}$$

In the above formula, D represents the density of the network, L is the number of existing edges, and g is the number of nodes in the graph.

# Result

## Theoretical Analysis of Rights and Obligations Relationship Diagram

In terms of the current research status within the academic community, in the field of analyzing the rights and obligations of credit card service contracts, some scholars such as Xu Yun (2023) and Chen Yue (2024) typically categorize the legal relationship between card issuers and cardholders into four types: savings and loan relationship, consumer credit relationship, entrustment and agency settlement relationship, and guarantee relationship, and they provide a brief analysis of the basic implications of each relationship. Among these, the consumer credit relationship, where the cardholder is required to repay the principal and interest on time after overdrawing, is most closely aligned with the focus of this study. However, it is regrettable that the aforementioned literature does not explicitly define the rights and obligations of each party, and there is a lack of in-depth analysis on the rights and obligations of all parties in the existing literature. In terms of personal information protection, Xie Ni (2020) pointed out that the standardized terms in credit card contracts pose potential risks that could endanger the security of personal information when dealing with the use and storage of such information; Chang Liying (2023) also noted that the current credit card debt collection model carries risks that could lead to the leakage of users' personal information.

Synthesizing the above research findings and the Supreme People's Court's "Provisions on Several Issues Concerning the Trial of Civil Dispute Cases Involving Bank Cards" (hereinafter referred to as the "Provisions")[3], in the area of credit card disputes involving personal information protection, the rights and obligations of each party are relatively clear and suitable for transformation into a rights relationship diagram. Specifically, in the credit card service contract, the cardholder has the obligation to repay the overdraft principal and interest on time and enjoys the rights to consumer credit and to supervise the debt collection behavior; the card issuer has the right to entrust and supervise external collection agencies to collect overdue debts and simultaneously has the obligation to protect the cardholder's information security; the external collection agency is responsible for collecting overdue debts, must regularly report progress to the entrusting bank, and is obligated to protect the cardholder's privacy. Based on the above, this paper has constructed a social relationship network diagram based on rights and obligations. (See Figure 4)

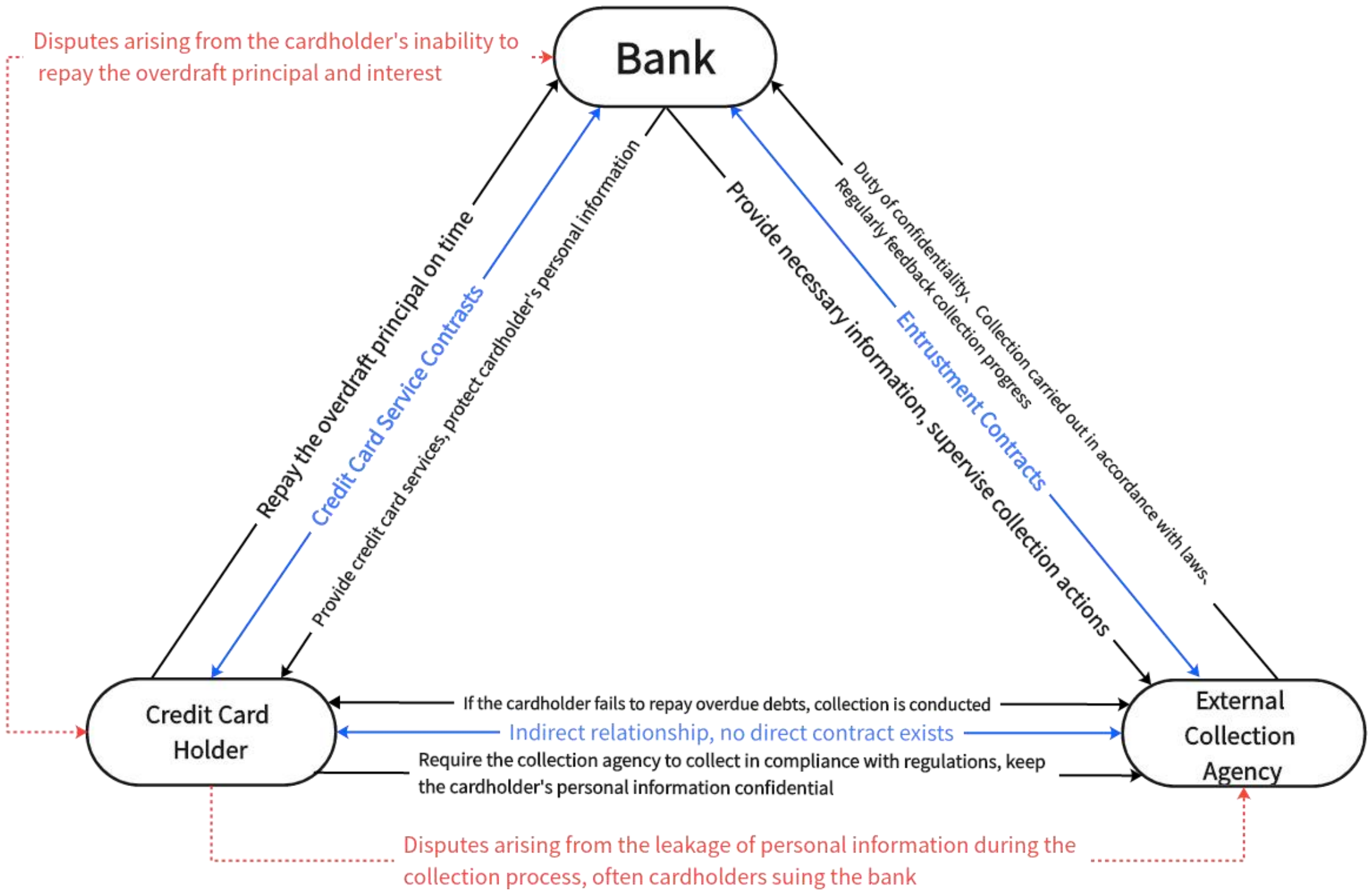

Figure 4

# Result Analysis

1. Building a Credit Card Contract Dispute Risk Network Based on the Regularities of Legal and Regulatory Citations

In the legal and regulatory citation network, with relevant laws and regulations as the rows and 48 judgement documents as the columns, a membership matrix (affiliation matrix) is jointly established. The element Xij in the matrix indicates whether the i-th law or regulation is cited in the j-th judgement document. If it is cited, then Xij=1 (true); if not, Xij=0 (false), thereby constructing a “laws and regulations - judgement documents” two-mode relationship matrix. For this legal citation structure matrix, the UCINET software is utilized to build the legal citation network graph G (N, K), where N represents the number of nodes in the network, and K represents the number of network connections with weighted values. The visualization of this network is presented in Figure 5.

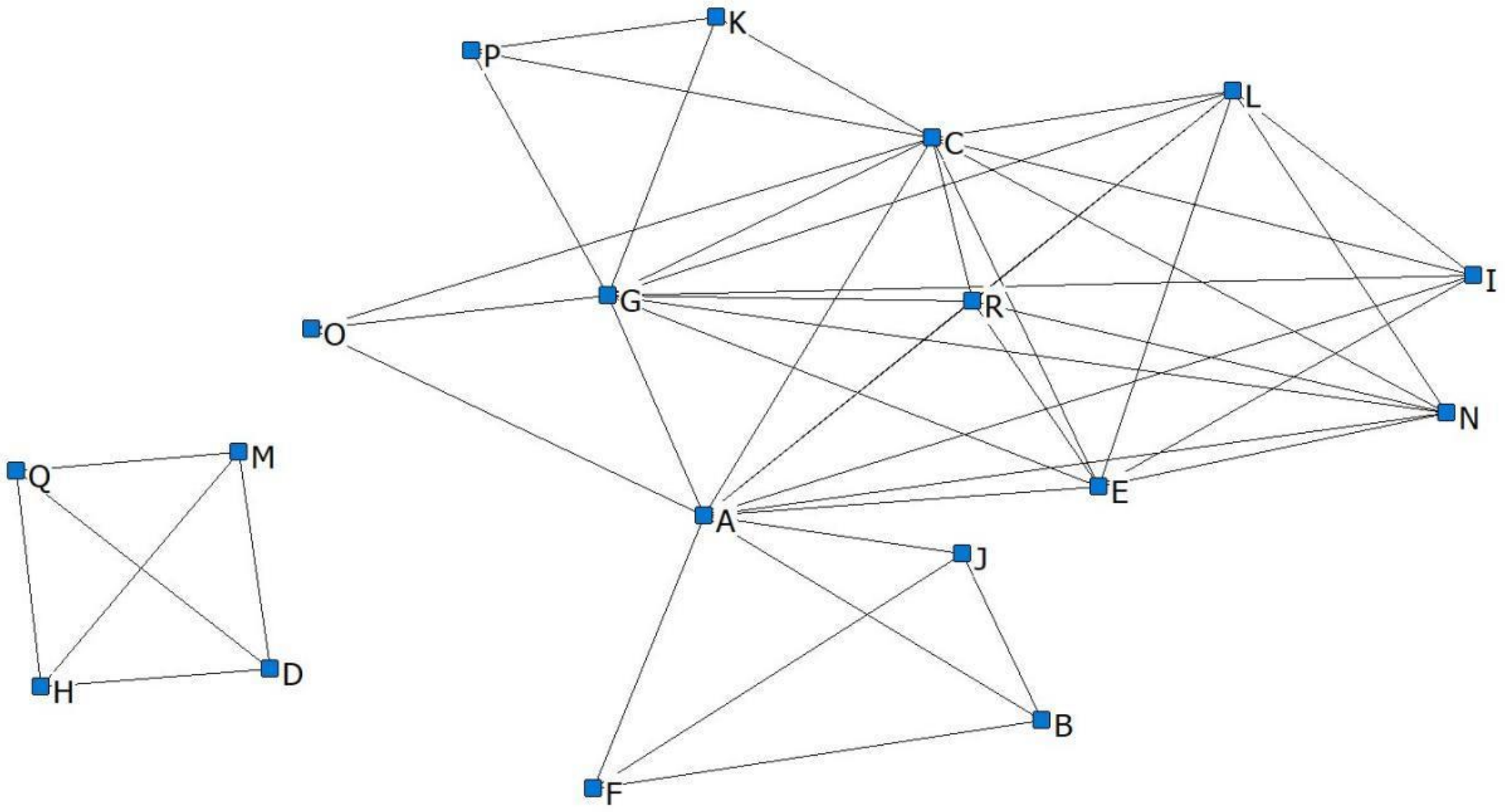

Figure 5

In Figure 5, the quadrilateral QMDH forms a separate shape and is not connected to any other nodes (i.e., laws and regulations). Upon analysis, all four items belong to the same case, which is "Civil Judgement of First Instance in the Credit Card Dispute Case between Beijing Shunyi Branch of Agricultural Bank of China Limited and Jin Yong." The authors consider that the cause of action in this case significantly differs from and is more complex than other cases. Therefore, this case was excluded from the subsequent analysis in the research process.

| **Legal Provisions** | **Serial Number** | **Degree** | **Betweenness Centrality** |
|---|---|---|---|
| Several Provisions of the Supreme People's Court on the Retroactivity in the Application of the CivilCode of the People's Republic of China,Art.1 | A | 11 | 32.067 |
| Civil Code of the People's Republic of China Art.1032 | B | 3 | 0 |
| Contract Law of the People's Republic of China,Art.60(Invalidated by Civil Code of the People's Republic of China,Book Three Contracts,Art.509) | C | 10 | 12.067 |
| Civil Code of the People's Republic of China,Art.496 | D | 3 | 0 |
| Civil Code of the People's Republic of | E | 7 | 0.4 |
| China,Art.6 | | | |
| Civil Code of the People's Republic of China,Art.1034 | F | 3 | 0 |
| Contract Law of the People's Republic of China,Art.107(Invalidated by Civil Code of the People's Republic ofChina,Book Seven Tort Liability,Art.577) | G | 10 | 12.067 |
| Civil Code of the People's Republic of China,Art.497 | H | 3 | 0 |
| Contract Law of the People's Republic of China,Art.8(Invalidated by Civil Code of the People's<br>Republic of China,Book Three Contracts,Art.465) | I | 5 | 0 |
| Civil Code of the People's Republic of China,Art.1035 | J | 3 | 0 |
| The Guarantee Law of the People's Republic of China,Art.18(Invalidated by Civil Code of the People's Republic of China,Book Three Contracts,Art.688) | K | 3 | 0 |
| Provisions of the Supreme People's Court on Several Issues Concerning the Trial of Cases Regarding Civil Disputes over Bank Cards,Art.2 | L | 7 | 0.4 |
| the Measures for the Administration ofBanke Card Business,Art.6 | M | 3 | 0 |
| Civil Procedure Law of the People's Republic of China,Art.147 | N | 6 | 0 |
| Civil Procedure Law of the People's Republic of China,Art.144 | O | 3 | 0 |
| Several Provisions of the Supreme People's Court on the Retroactivity in the Application of the Civil Code of the People's Republic of China,Art.20 | P | 3 | 0 |
| Measures for the Supervision and Administration of the Credit Card Business of Commercial<br><br>Banks,Art.39 | Q | 3 | 0 |
| Civil Procedure Law of the People's Republic of China,Art.95 | R | 6 | 0 |

Table 1 Degree and Betweenness Centrality of Nodes Related to Laws and Regulations

(Compiled and Illustrated by the Author)

2. Calculation of Network Indicators for Credit Card Contract Disputes Based on

the Patterns of Quoted Laws and Regulations:

Using UCINET software to measure the individual structure indicators of the legal and regulatory citation network, the degree and betweenness centrality of the nodes are calculated, which are shown in Table 1.

At the same time, based on the overall network-related indicators, the calculations are as follows in Table 2:

| | |
|---|---|
| **Density** | **0.301** |
| **Number of Edges** | **92** |
| **Number of Nodes** | **18** |

Table 2 Overall Metrics (Compiled and Illustrated by the Author)

When analyzing network density, Wellman believes that a density ranging from 0 to 0.25 indicates a lower relational density, meaning the connections are sparse (Wellman B., 1979). Based on this, it can be inferred that the legal and regulatory network of credit card contract disputes in Beijing from 2022 to 2024 has relatively dense connections, implying a significant degree of association among various laws and regulations. From this, it can be deduced that the commonalities in the logic of adjudication inferred from this network have a high degree of credibility.

## Analysis and Interpretation of Legal Relationship Network Model

The legal relationship network model constructed based on the social relationship network model analyses and correlates the effective judgements of a large number of similar cases. Using this network, judges in digital courts can refer to relevant legal provisions when handling cases, aiding in case adjudication. This is how the "digital assistance in case handling" function of digital courts can be realised. The criteria for identifying similar cases are based on the three elements of "essential facts," "dispute focus," and "application of law." From the perspective of "application of law," there are issues in the results of the legal relationship network model that require further discussion and clarification. If not studied and interpreted, they may affect the reference and judgement of judges in smart courts. Therefore, in order to better assist judges in digital courts in trying similar cases, this paper intends to conduct a preliminary analysis of the results of the legal relationship network model formed by effective judgements before analysing the overall role of the model.

In the legal relationship network, there is a positive correlation between degree and betweenness centrality. As previously mentioned, nodes with higher degrees and betweenness centrality have greater control over other laws and are located closer to the center of the network. Nodes A, C, and G have the highest degrees and betweenness centrality, with their frequency in the legal relationship network reaching or approaching 50%, indicating that these provisions are a major component of the trial logic in the network.

In the social relationship network formed by the laws and regulations cited in personal information-related bank card disputes, the connections between nodes are relatively close, meaning that laws and regulations often appear together in the same judgement. At the same time, nodes A, C, and G have significantly higher degrees and betweenness centrality, indicating they occupy a more central position in the logic of this type of judgement. Meanwhile, this paper finds that the establishment of the network graph can quickly exclude cases that do not match the type. For example, after establishing a network graph based on credit card contract disputes, nodes that have no connection to the overall network (such as the quadrilateral QMDH in Figure 3) were found, upon confirmation, to have causes of action that significantly differ from other cases, and can thus be quickly filtered out from the database of similar cases. In other words, the use of the network provides a possible path for the rapid classification of case types.

Firstly, credit card disputes involving personal information protection exhibit characteristics of typed distribution. Previous research has shown that in the legal relationship network, the network density indicator is greater than 0.3, which means that the cited laws and regulations show a clustered distribution with relatively close connections. In other words, the judgement logic of the cases studied in this paper has strong commonalities, forming a typification of judgement logic. Based on this, it can be preliminarily determined that specific credit card contract dispute cases in Beijing have a high degree of consistency in judgements, and therefore, they are suitable for establishing a database for searching similar cases. This database has strong universality and can form systematic judgement guidance for similar disputes.

Secondly, the provisions on the temporal validity of the application of the "Civil Code" are at the center of the legal relationship network. In the network, node A has the highest degree and betweenness centrality, with the "degree" reaching 12 and the "betweenness centrality" exceeding 32. In other words, most of the effective judgements studied in this paper have cited this article. Based on this, the paper finds that the issue of the statute of limitations is related to the personalized installment repayment agreement. In 2011, Article 70 of the "Commercial Bank Credit Card Business Supervision and Management Measures" proposed provisions on personalized installment repayment agreements, which allow card-issuing banks to enter into such agreements with cardholders when the cardholder is unable to repay the credit card debt and has the willingness to repay. In 2022, the China Banking and Insurance Regulatory Commission and the People's Bank of China supplemented and reaffirmed this provision: prohibiting the re-subscription of installment payments for funds that have already been processed in installments, clarifying that the longest term for installment business is 5 years, and strengthening the regulation of credit card installment business. Based on this, credit cardholders may have been unable to repay the overdrawn credit card before the implementation of the "Civil Code," but the existence of the personalized installment repayment agreement has prolonged the litigation activities until 2022 or 2023, i.e., after the "Civil Code" came into effect. Therefore, the extensive citation of the provisions on the temporal validity of the application of the "Civil Code" can be explained; moreover, although the original "Contract Law" has been invalidated since 2021, its provisions are still frequently cited in judgements. The reason for this phenomenon can also be reasonably explained based on the analysis of this section.

Thirdly, the general provisions in the original “Contract Law” occupy an important position in the legal relationship network. Data from the network show that Article 60 and Article 107 of the original “Contract Law” (corresponding to nodes C and G in the grid diagram) have significantly higher degrees and betweenness centrality, and these two regulations are frequently cited together in multiple cases. In contrast, the “Supreme People’s Court’s Several Provisions on Trying Civil Disputes over Bank Cards” (hereinafter referred to as the “Provisions”) have lower degrees and betweenness centrality. The paper’s analysis of the provisions of the original “Contract Law” finds that Article 60 stipulates the principle of comprehensive performance of the contract and the ancillary obligations of performance, while Article 107 stipulates the basic forms of breach of contract and the types of liability for breach of contract. At the same time, the “Provisions” were formulated in the context of the rapid development of Internet finance to address the increasing number of disputes over bank card fraud, overdrawn interest and fees, and penalty charges. Combining the legal obligation relationship in credit card disputes involving personal information protection, the paper finds that the rights and obligations in such disputes are relatively fixed, and the causes of action mainly include the following two: first, the cardholder’s failure to fulfill the obligation to repay the principal and interest on time; second, the leakage of the cardholder’s personal information. The former involves the cardholder’s obligation to pay, which is a primary obligation stipulated in the contract. The latter is the obligation to protect the cardholder’s personal information, which is a legal ancillary obligation. Article 60 of the original “Contract Law” prohibits a party to the contract from failing to perform the primary obligation and ancillary obligations, and if violated, the party shall bear the liability for breach of contract as stipulated in Article 107 of the original “Contract Law.”

Credit card disputes involve multiple types, including fraud, overdrawn interest and fees, and penalty charges. The credit card disputes involving personal information protection studied in this paper are only one of them. Considering the background and content of the “Provisions,” it can be seen that the “Provisions” have little relation to the type of credit card disputes studied in this paper, and the content of the original “Contract Law” is more suitable for resolving such disputes. Therefore, in judicial practice, more provisions of the original “Contract Law” are cited rather than the clauses of the “Provisions.”

| Laws and Regulations | Introduced Time | Related Contend |
|---|---|---|
| Several Provisions of the Supreme People’s Court on the Retroactivity in the Application of the Civil Code of the People’s Republic of China, Art.1 | 2020-12-29 | General provisions on the time effect of the Civil Code |
| Civil Code of the People’s Republic of China Art.1032 | 2021-01-01 | Provisions relating to privacy and the right of privacy |
| Contract Law of the People’s Republic of China, Art.60（Invalidated by Civil Code of the People’s Republic of China, Book Three Contracts, Art.509） | 1999-10-01 | Fully perform its own obligations as agreed and the principle of good faith. |
| Civil Code of the People’s Republic of China, Art.496 | 2021-01-01 | Provisions relating to standard term. |
| Civil Code of the People’s Republic of China, Art.6 | 2021-01-01 | The principle of fairness |
| Civil Code of the People’s Republic of China, Art.1034 | 2021-01-01 | The personal information of natural persons is protected by law. |
| Contract Law of the People’s Republic of China, Art.107（Invalidated by Civil Code of the People’s Republic of China, | 1999-10-01 | Provisions relating to the liability for breach of contract. |
| Book Seven Tort Liability, Art.577 ） | | |
| Civil Code of the People’s Republic of China, Art.497 | 2021-01-01 | Voidable standard terms |
| Contract Law of the People’s Republic of China, Art.8（Invalidated by Civil Code of the People’s Republic of China, Book Three Contracts, Art.465） | 1999-10-01 | Principle of performing the obligations in accordance with the terms of the contract. |
| Civil Code of the People’s Republic of China, Art.1035 | 2021-01-01 | The personal information of a natural person shall be processed under the principles of lawfulness, justification and necessity. |
| The Guarantee Law of the People’s Republic of China, Art.18（Invalidated by Civil Code of the People’s Republic of China, Book Three Contracts, Art.688） | 1995-10-01 | Provisions relating to the jointly and severally liable for the obligation. |
| Provisions of the Supreme People’s Court on Several Issues Concerning the Trial of Cases Regarding Civil Disputes over Bank Cards, Art.2 | 2018-06-06 | The bending force of the term of the contract on bank card and the principle of ruling. |
| the Measures for the Administration of Banke Card Business, Art.6 | 1999-01-18 | The classification of the credit cards. |
| Civil Procedure Law of the People’s Republic of China, Art.147 | 2017-06-27 | Provisions relating to the default judgement. |
| Civil Procedure Law of the People’s Republic of China, Art.144 | 2017-06-27 | The order of the court debate. |
| Several Provisions of the Supreme People’s Court on the Retroactivity in the | 2020-12-29 | Provisions on the temporal effect of the Civil Code on |
| Application of the Civil Code of the People's Republic of China, Art.20 | | cross-jurisdictional contract. |
| Measures for the Supervision and Administration of the Credit Card Business of Commercial Banks, Art.39 | 2011-01-13 | Marketing management system of credit cards. |
| Civil Procedure Law of the People's Republic of China, Art.95 | 2017-06-27 | Provisions relating to the service of process by public announcement. |

Table 3 Three-line table of laws and regulations (Compiled and Illustrated by the Author)

# Discussion

As Max Weber and Montesquieu's "vending machine" theory posits, "modern judges are like vending machines. People insert lawsuits and litigation fees, and what comes out are judgements and reasons copied from the code (Weber M., 1998 )." This theory emphasizes ensuring that the law maintains the original intent of the legislator in its application and prevents application arbitrarily. Given the increasingly prominent problem of chaotic legal norms and the lack of uniformity in China's legal system, to ensure the uniformity of the legal system and achieve the goal of justice it is not sufficient to have judges across the country apply the same legal provisions when making judgements, for that the same legal provisions and legal concepts can be interpreted differently (He WF., 2003). Therefore, our country needs a "vending machine-style" intelligent trial system that can eliminate external interference. With this system, one only needs to "input" the course of the case and the focus of the dispute. After "analysis" of laws and regulations and past precedents, the system will quickly "output" the judgement results and relevant legal basis, assisting judges in making judgements.

Integration of artificial intelligence and trial procedures occurred both internationally and domestically. Internationally, International Business Machines Corporation has launched the Advanced Case Management judicial assistance system. Domestically, the Nanjing

Intermediate People's Court has introduced the "Alpha Judge" robot-assisted case-judgement system (Sheng XJ., Zou Y.， 2018) However, current practice has only focused the construction of digital courts more on the technical perspective rather than the legal one, which is the reason why the development of smart courts struggles to reach a higher level (Xu YY., Zhu YN., 2020). Moreover, an intelligent system with powerful storage, search, computing, and learning functions may not necessarily become a qualified "robot judge". It should possess the concepts of human nature and humanity, believe in the spirit of justice, and understand the art of adjudication to settle disputes (Sheng XJ., Zou Y., 2018).

Therefore, the study aims to reconstruct the smart trial system around the key point of a "digital holistic system", using big data and artificial intelligence to process judicial data, assisting judges in quickly retrieving information while thinking independently and achieving the goal of rapidly processing simple cases and more efficiently judging major and difficult cases, ultimately achieving justice.

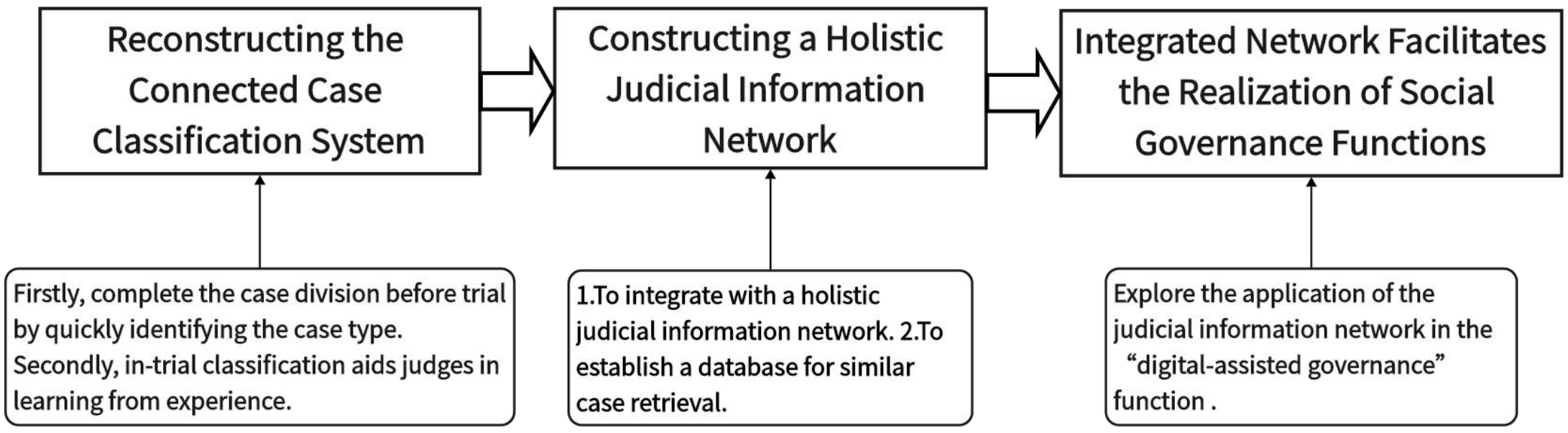

Figure 6

**(I) Reconstructing the Connected Case Classification System to Achieve Justice**

To achieve the goal of building a holistic system, the first step is to sort out and classify the existing massive amount of data, which will facilitate subsequent retrieval, analysis, and network construction. This first step focuses on determining reasonable classification methods and principles that balance classification accuracy with convenience.

Firstly, complete the case division before trial by quickly identifying the case type.

In traditional judicial case databases, case retrieval and case analysis are separate processes. Judges must first classify and retrieve cases based on the cause of action in the database, and then analyse the existing cases within the type of dispute before finalizing the classification.

This procedure is cumbersome and reduces the efficiency of trials to a certain extent. This study firstly analyses the rights relationships and dispute resolution pathways. Then, with the help of social relationship network tools, this study analyses the cited laws and regulations in these cases and constructs the overall network.

This finding proves to a certain extent that the methodology of typifying cases and analyzing their characteristics and commonalities can be extended to all cases in the process of building “digital courts”. By leveraging artificial intelligence semantic understanding and training engines, this study can extract the cases needed by judges through keyword retrieval in the massive judicial case library, classify them, and automatically generate information of typified cases. This means that judges only need to enter case keywords into the system, and the smart trial system can automatically identify the case type based on the keywords. It then provides a database of the type of case and the common issues encountered when hearing such cases. This significantly streamlines trial procedures, reduces the workload of judges, and thereby improves trial efficiency, contributes to the goal of justice.

Secondly, in-trial classification aids judges in learning from experience.

Additionally, current case databases are largely classified by the cause of the case, and cases within a category can only be sorted by external features such as time, jurisdiction level, and document type in ascending or descending order. However, in trial practices, classification by cause can be quickly achieved based on experience, and judges require a more detailed classification case database. Nowadays, artificial intelligence is not only unable to detach itself from the established knowledge and practices of humans but also has a deep dependence on human existence (Huang DD., 2020), which made most current classification systems mechanized and “foolproof”, and it determines the category of a case solely based on the presence of certain keywords, which is inappropriate. For instance, during the search process for this article, many cases were found to be categorized as personal information cases simply due to the submission of a personal information registration form in the judgement document.

In the process of network construction for 49 cases of this dispute type, this study found that the laws and regulations cited in one case had no connection with the judicial regulations cited in other cases. Through analysis and comparison of this case, this study chose to exclude this case from discussion. The process of exclusion mentioned above can only be achieved manually at present, requiring judicial officers to categorize with the help of social networks. However, if the artificial intelligence system is trained by judicial officers through generating data on the classification and exclusion process, it is hoped that the current classification system can be improved, and a more professional and accurate database, enriched with the wisdom of judicial officers' experience, can be constructed.

Through the construction of a keyword search and classification system and the reconstruction of a precise classification system, it can address the issues of mechanization and superficiality present in the current classification system, and enhance the efficiency of trials from case classification to judgement, helping achieve justice.

**(II) Constructing a Holistic Judicial Information Network to Achieve Justice**

After rebuilding the classification system and clustering individual cases, the next step is to identify linkages within and between types of disputes to build a holistic network of cases, and then construct a res judicata analysis model.

The first goal is to integrate with a holistic judicial information network. In judicial trials, every case contains an information chain (Jia Y., 2024). If only looking at individual information points in isolation, it is difficult to discover the connections between cases and the specific issues of this type of case and hinder the achievement of justice. However, cases involved in traditional legal research exist as individual instances, and judges can only analyse the cases to be tried one by one and independently, without being able to adequately compare each case with big data. This leads to a lack of intersection and correlation between case data, making it difficult to clearly show the commonality and individuality that exists in many cases, and it is also impossible to deeply explore and summarize the rules of experience from the increasing number of judgement cases.

Within typified disputes, this study takes the laws and regulations cited in the judgements as the logical vein, establishes a judicial information data network model, constructs multiple isolated judgement 'samples' in the typified disputes into a clear visual information network, and generates the number of times that each law and regulation is cited and the strength of the relationship between the laws and regulations, thus realizing the interactive linkage of res judicata. In judicial practice, judges can refer to the applicable network laws and regulations of this typified dispute during the trial process. Meanwhile, the court can analyse and compare the trials of different judges on the same typified disputes, test whether the judges have preferences for the way of judgement, avoid the situation of "different judgements for the same case," maintain the judicial res judicata, and strengthen the quality supervision and management of cases and ultimately achieve the goal of justice.

Additionally, the court can test the implementation and enforcement effects of relevant laws and regulations by generating a network of laws and regulations in a specific time and historical context. For example, this study found that in the social network model, the clauses on time effectiveness were cited many times. It can be seen that a large number of cases were instituted before the promulgation of the Civil Code and were judged after the promulgation of the Civil Code, involving disputes about the applicable law.

The second goal is to establish a database for similar case retrieval. China's Code of Civil Procedure still has some important aspects that need enhancement or even missing, such as the judgement system. The current system is overly simplistic, lacking clear provisions on objectives, subjectivity, time limits of res judicata, and the relationship between the main text of the judgement (Zheng K., 2021). This has impacted the unification of our country's legal system and judicial credibility. Therefore, it is essential to establish a mechanism for higher courts to coordinate the legal decisions of lower courts uniformly, ensuring consistency in legal interpretation and achieving procedure justice.

This study aims to promote a searchable database of classified cases through the establishment of a judicial information network, forming a network of res judicata through precedents and the judicial interpretations and guiding cases of the People's Supreme Court. This network will bind the judges' decision-making power, preventing miscarriages of justice in the form of "different judgements for the same case", ensuring that once a final judgement is determined, the parties cannot dispute the decision, nor can the court make a new judgement that contradicts it. The res judicata judgement has a mandatory binding effect on subsequent trials. By constructing an intelligent trial system based on the case retrieval database, we can move away from the traditional trial mode of dividing cases according to the type of cause of action, and instead tap into the deeper linkages between cases. This is conducive to strengthening trial supervision, enhancing judicial credibility, and achieving the goal of justice.

**(III) Potential Risks in Achieving Justice**

Certain risks exist regarding the application of AI and big data in procedural and substantial matters of judicial adjudication, both in technical and ethical aspects (Zhang Q., Chen S., Song T., 2023). Technically, incomplete algorithms may delay the progress of the trial and hinder the achievement of justice. Ethically, the question of who is responsible for the wrong judgements made by AI has also been a long-debated issue (ibid.). This study argues that AI and big data merely serve as tools to assist human judges in making decisions, and agrees that to realise the goal of justice, human judges must make and supervise the final judgement (Shi, J., 2022). The technical shortcomings of the algorithm need to be refined in subsequent research and practice, underlying debate on ethics also remains to be solved in a long time.

# Limitations and Future Studies

Applying quantitative analysis tools and social relationship networks to digital courts can produce more intuitive and feasible results, such as judicial data networks and case-type retrieval databases. Hothis studyver, this study has some limitations. Firstly, it focuses solely on credit card disputes involving personal information protection in Beijing, and the replicability of the legal and regulatory system derived from our analysis in other regions requires further verification. Secondly, this study has chosen the PKU Law Database as our primary source for case retrieval, which may limit the comprehensiveness of case coverage. Thirdly, this study concentrates on the laws and regulations relevant to case judgements and does not analyse other keyword networks that could be constructed using social network analysis methods.

This study aims to explore the precision of law application, support trials and ultimately achieve justice. In future studies, this paper will build a case data network in complex cases by abstracting the complex case merits into a fusion of different simple case models, and will explore the establishment of networks with different

keywords, which would link cases from various perspectives, such as time and geography, to assist judges in conducting case analyses and summarizing their experiences from multiple entry points, both horizontally and vertically. This approach is hoped to contribute to the future realization of a holistic system of judicial information.

# Conclusion

This study examines typical disputes by analysing effective judgements from relevant cases in Beijing from 2022 to 2024, creating a database of cited regulations. It contributes to the field by innovatively applying quantitative analysis tools and social relationship analysis in digital courts. The research establishes the feasibility of using judicial data networks and case-type retrieval databases in developing digital courts and proposes a three-step strategy—case clustering, network linking, and service enhancement—to improve trial processes. These measures offer effective solutions to the challenges and shortcomings of traditional case assignment systems in courts, contributing to the pursuit of justice.

# Data availability

To ensure the reliability and consistency of the research results, this paper employs Cronbach's α reliability coefficient for reliability testing and uses the SPSSAU software for calculation. The results are presented in Table 4.

| Name | Corrected Item-Total Correlation (CITC) | Cronbach's a ifItem Deleted | Cronbach's α |
|---|---|---|---|
| Item 1 | 0.758 | - | 0.586 |
| Item 2 | 0.758 | - | 0.586 |
| | Note:Standaerdized Cronbach's α=0.863 | | |

Table 4 Cronbach's Alpha Reliability Analysis Results

Table 4 shows that the Cronbach's α coefficient for BA, PC, and SN is 0.586,

which is greater than 0.5 but less than 0.6. However, it consists of only two analysis items, indicating that the quality of research data reliability is acceptable. Regarding the “CITC value,” the CITC values of the analysis items are all greater than 0.5, suggesting a good correlation between the analysis items and also indicating a good level of reliability, demonstrating that the variables have good internal consistency reliability. In summary, the research data used in this paper has a certain degree of reliability and can be used as a basis for further research.

# Competing interests

The author(s) declare no competing interests.

# Ethical statements

This article does not contain any studies with human participants performed by any of the authors.

# Numbered Endnotes

[1] Judgment documents related to all adjudicated procedures (including first-instance, second-instance, and retrial) by the Beijing Higher People's Court, Intermediate People's Courts, and Primary People's Courts in various jurisdictions.

[2] Civil Procedure Law of the People's Republic of China, Art.160: Where a primary people's court and its detached tribunals try simple civil cases with clear facts, unambiguous rights and obligations and minor disputes, the provisions of this Chapter shall apply.

Where a primary people's court and its detached tribunals try civil cases other than those in the preceding paragraph, the parties may agree on the application of summary procedure.

[3] Adopted at the 1785th meeting of the Judicial Committee of the Supreme People's Court on December 2, 2019, and effective from May 25, 2021.

# Figure legend

| Figure Number | Name of the Contents |
| --- | --- |
| Figure 1 | Number and percentage of cases applying summary proceedings nationwide (compiled and self-illustrated herein) |
| Figure 2 | Number and percentage of cases applying summary procedures in Beijing (collated and self-collected herein) |
| Figure 3 | Overall Structure of the Paper (Compiled and Illustrated by the Author) |
| Figure 4 | Based on the Schematic Diagram of Credit Card Rights and Obligations Relationship (Compiled and Illustrated by the Author) |
| Figure 5 | Beijing Municipal People's Court and Its Subordinate Courts' Legal and Regulatory Citation Network in Adjudicating Credit Card Contract Dispute Cases (Compiled and Illustrated by the Author) |
| Figure 6 | Digital Holistic System Construction Roadmap |